\documentclass [a4paper,fleqn,12pt]{article}
\usepackage{graphicx}
\usepackage[small]{subfigure,epsfig}
\usepackage{indentfirst}
\usepackage {amsmath} \usepackage{amssymb} \usepackage{cite}

\begin{document}

\title{Comment on: ''New exact solutions for the Kawahara equation using Exp-function method'' }

\author{Nikolai A. Kudryashov\footnote{E-mail: kudryashov@mephi.ru}}

\date{Department of Applied Mathematics, National Research Nuclear University
''MEPhI", 31 Kashirskoe Shosse, 115409 Moscow, Russian Federation}

\maketitle

\begin{abstract}

Exact solutions of the Kawahara equation by  Assas [L.M.B. Assas, J. Com. Appl. Math.
233 (2009) 97--102] are analyzed. It is shown that all solutions do not satisfy the
Kawahara equation and consequently all nontrivial solutions by Assas are wrong.
\end{abstract}



Recently Assas in \cite{Assas_2009} looked for exact solutions of the Kawahara equation using
the Exp - function method. This equation can be written as
\begin{equation}\label{Eq1}
u_t+\alpha\,u\,u_x+\beta\,u_{xxx}-\gamma\,u_{xxxxx}=0,
\end{equation}
where $\alpha$, $\beta$ and $\gamma$ are constants.

The author \cite{Assas_2009} considered the Kawahara equation using the traveling wave
$z=k\,x+w\,t$ and  tried to find new exact solutions of the nonlinear ordinary differential equation
\begin{equation}\label{Eq1a}
\gamma\,k^5\,U_{zzzzz}-\beta\,k^3\,U_{zzz}-\alpha\,k\,U\,U_{z}-w\,U_{z}=0.
\end{equation}

He believe that he found a few exact solutions of Eq.\eqref{Eq1a} but this is not the case.  It is obviously that the author \cite{Assas_2009} can not obtain any new  nontrivial solutions of Eq.\eqref{Eq1a} using his approach. The matter is the general solution of Eq.\eqref{Eq1a} has the pole of the fourth order.  This fact  has an important bearing on the choose of expressions in the Exp-function method. However the author does not take this result into account. To look for exact solutions of the Kawahara equation the author \cite{Assas_2009} must take at least five terms in denominator and numerator of the ansatz in the Exp-function method. We can see that all constructions by author (formulae (15), (21) and (27) of work \cite{Assas_2009}) have poles of the second and the third order.

Consequently we can expect that Assas can not find any solutions of the Kawahara equation. Just to be on the save side we have checked all solutions by Assas \cite{Assas_2009} and obtained that all solutions \cite{Assas_2009} do not satisfy the Kawahara equation except solution (17) that is the trivial solution $u=\frac{a_{-1}}{b_1}$.

We were surprised when we read that the author \cite{Assas_2009} has compared the accuracy his exact solutions with known exact solution and obtained that his results are "very precise and that there is an inverse relationship between distance and time". In conclusions the author claim that "some new generalized solitary solutions with parameters are obtained".

In fact solitary and periodic solutions of the Kawahara equation were obtained many years ago  (see for example \cite{Kudr_88, Kudr_90, Kudr_90a}). Let us demonstrate that we can find the solitary wave solutions of the Kawahara equation using the tanh - method. Assuming $U(z)$ in the form
\begin{equation}\begin{gathered}
\label{Eq2}U(z)=A_0+A_1\,\tanh \left( {z} \right) +A_{{2}} \tanh^2
 \left( {z} \right)  +\\
\\
 +A_{{3}} \,\tanh^3 \left( {z} \right)  + A_{{4}} \tanh^4 \left( {z} \right), \qquad z=k\,x+ w\,t-z_0
  \end{gathered}\end{equation}
and substituting \eqref{Eq2} into Eq.\eqref{Eq1a} we have
\begin{equation}\begin{gathered}
\label{Eq3}A_{{4}}=\,{\frac {1680\,\gamma\,k^4}{\alpha\,}}, \quad A_{3}=0,\qquad A_{{2}}=- \,{\frac {280\,k^2\,(104\,\gamma\,k^2+\beta)}{13\,\alpha\,}},\qquad A_{1}=0, \\
\\
A_{{0}}=\,{\frac {264992\,{\gamma}^{2}\,k^5+7280\,\beta\,\gamma\,k^3-31\,{
\beta}^{2}\,k-507\,\gamma\,w}{507\,\gamma\,\alpha\,k}}
  \end{gathered}\end{equation}

As result we obtain the solitary wave solutions in the form
\begin{equation}\begin{gathered}
\label{Eq2a}u(x,t)=\,{\frac {264992\,{\gamma}^{2}\,k^5+7280\,\beta\,\gamma\,k^3-31\,{
\beta}^{2}\,k-507\,\gamma\,w}{507\,\gamma\,\alpha\,k}} - \\
\\-  \,{\frac {280\,k^2\,(104\,\gamma\,k^2+\beta)}{13\,\alpha\,}}\, \tanh^2
 \left( z \right)
 +\,{\frac {1680\,\gamma\,k^4}{\alpha\,}}\, \tanh^4 \left( z \right), \\
\\
 z=k\,x+ w\,t-z_0
  \end{gathered}\end{equation}
 with the following values of $k$:
\begin{equation}\begin{gathered} {k_{1,2}}=\pm\,{\frac {\sqrt {13\,\gamma\beta}}{26\,\gamma}},\qquad k_{3,4}=\pm\,{\frac {\sqrt {65\,\gamma\beta\, \left( -31+3\,i
\sqrt {31} \right) }}{260\,\gamma}},\\
\\
k_{5,6}=\pm\,{\frac {\sqrt {-65\,\gamma\beta\, \left( 31+3\,i\sqrt {31
} \right) }}{260\,\gamma}}.
 \end{gathered}\end{equation}

Other solutions in the form of the solitary waves of the Kawahara equation are not known. More then that we are sure that we can not find other solitary wave solutions of the Kawahara equation because there are meromorphic solutions only in these cases. Unfortunately the author \cite{Assas_2009} made a few mistakes that were discussed in recent papers \cite{Kudr_2009a, Kudr_2009b, Kudr_2009c, Kudr_2009d, Parkes}.

\textbf{Conclusion. } This comment was sent in september 2009 to the Journal of Computational and Applied Mathematics. It was the long story with this comment. However recently I received the decision from the Principal Editor  M.J. Goovaerts: "The paper on which you are giving comments will be retracted". Certainly It will be nice solution on wrong paper but I believe it is not easy task. Let us wait. Maybe it will happen.

\end{document}